\documentclass[12pt]{article}

\usepackage{epsf}
\usepackage{rotate}

\def\spose#1{\hbox to 0pt{#1\hss}}
\def\lsim{\mathrel{\spose{\lower 3pt\hbox{$\mathchar"218$}}
 \raise 2.0pt\hbox{$\mathchar"13C$}}}
\def\gsim{\mathrel{\spose{\lower 3pt\hbox{$\mathchar"218$}}
 \raise 2.0pt\hbox{$\mathchar"13E$}}}

\begin{document}

\begin{titlepage}

\begin{flushright}
CERN-TH/99-78\\
hep-ph/9903455
\end{flushright}

\vspace{2cm}
\begin{center}
\boldmath
\large\bf
Extracting $\gamma$ from $B_{s(d)}\to J/\psi\, K_{\rm S}$ and 
$B_{d (s)}\to D^{\,+}_{d(s)}\, D^{\,-}_{d(s)}$ 
\unboldmath
\end{center}

\vspace{1.2cm}
\begin{center}
Robert Fleischer\\[0.1cm]
{\sl Theory Division, CERN, CH-1211 Geneva 23, Switzerland}
\end{center}

\vspace{1.7cm}
\begin{abstract}
\vspace{0.2cm}\noindent
A completely general parametrization of the time-dependent decay rates
of the modes $B_s\to J/\psi\, K_{\rm S}$ and $B_d\to J/\psi\, K_{\rm S}$
is given, which are related to each other through the $U$-spin flavour
symmetry of strong interactions. Owing to the interference of current--current
and penguin processes, the $B_s\to J/\psi\, K_{\rm S}$ observables 
probe the angle $\gamma$ of the unitarity triangle. Using the $U$-spin 
symmetry, the overall normalization of the $B_s\to J/\psi\, K_{\rm S}$
rate can be fixed with the help of the CP-averaged $B_d\to J/\psi\, K_{\rm S}$
rate, providing a new strategy to determine $\gamma$. This extraction of
$\gamma$ is not affected by any final-state-interaction effects, and its 
theoretical accuracy is only limited by $U$-spin-breaking corrections. 
As a by-product, this strategy allows us to take into account also the 
penguin effects in the determination of $\beta$ from 
$B_d\to J/\psi\, K_{\rm S}$, which are presumably very small, and to 
predict the direct CP asymmetry arising in this mode. An analogous strategy 
is provided by the time-dependent $B_d\to D^+D^-$ rate, if its overall 
normalization is fixed through the CP-averaged $B_s\to D^+_sD^-_s$ rate. 
\end{abstract}

\vfill
\noindent
CERN-TH/99-78\\
March 1999

\end{titlepage}

\thispagestyle{empty}
\vbox{}
\newpage
 
\setcounter{page}{1}

\section{Introduction}\label{intro}
It is well known that the ``gold-plated'' mode $B_d\to J/\psi\, K_{\rm S}$
\cite{bisa} plays an outstanding role in the determination of $\sin(2\beta)$, 
where $\beta$ is one of the three angles $\alpha$, $\beta$ and $\gamma$ of 
the usual non-squashed unitarity triangle \cite{ut} of the 
Cabibbo--Kobayashi--Maskawa matrix (CKM matrix) \cite{ckm}. First 
attempts to measure $\sin(2\beta)$ in this way, which is one of the major 
goals of several $B$-physics experiments starting very soon, have recently 
been performed by the OPAL and CDF collaborations \cite{exp}.

In this paper, we will have a closer look at the general structure of the
$B_d\to J/\psi\, K_{\rm S}$ decay amplitude arising within the Standard
Model, and at the one of its $U$-spin counterpart $B_s\to J/\psi\, K_{\rm S}$.
The two decays are related to each other by interchanging all down and 
strange quarks, i.e.\ through the ``$U$-spin'' subgroup of the $SU(3)$ 
flavour symmetry of strong interactions. Whereas the weak phase factor 
$e^{i\gamma}$ enters in $B_d\to J/\psi\, K_{\rm S}$ in a strongly 
Cabibbo-suppressed way, this is not the case in $B_s\to J/\psi\, K_{\rm S}$.
Consequently, there may be sizeable CP-violating effects in this $B_s$
decay, which are due to the interference between current--current and 
penguin operator contributions. Interestingly, the time evolution of the  
$B_s\to J/\psi\, K_{\rm S}$ decay rate allows us to determine $\gamma$.
To this end, we have to employ the $U$-spin symmetry to fix the overall
normalization of $B_s\to J/\psi\, K_{\rm S}$ through the CP-averaged 
$B_d\to J/\psi\, K_{\rm S}$ rate. This new strategy to extract $\gamma$ is 
not affected by QCD or electroweak penguin effects -- it rather makes use 
of these topologies -- and does not rely on certain ``plausible'' dynamical
or model-dependent assumptions. Moreover, final-state-interaction effects 
are taken into account {\it by definition}, and do not lead to any problems. 
The theoretical accuracy is only limited by $U$-spin-breaking corrections. 
An analogous strategy is provided by the time-dependent $B_d\to D^+D^-$ rate, 
if its overall normalization is fixed through the CP-averaged 
$B_s\to D^+_sD^-_s$ rate, and if the $B^0_d$--$\overline{B^0_d}$ mixing 
phase, i.e.\ $2\beta$, is determined with the help of 
$B_d\to J/\psi\,K_{\rm S}$. 

In particular the determination of $\gamma$ is an important goal for 
future $B$-physics experiments. This angle should be measured in a variety 
of ways so as to check whether one consistently finds the same result. 
There are several methods to accomplish this task on the market \cite{revs}.
Since the $e^+e^-$ $B$-factories operating at the $\Upsilon(4S)$ resonance 
will not be in a position to explore $B_s$ decays, a strong emphasis has been 
given to decays of non-strange $B$ mesons in the recent literature. However, 
also the $B_s$ system provides interesting strategies to determine $\gamma$. 
In order to make use of these methods, dedicated $B$-physics experiments at 
hadron machines, such as LHCb, are the natural place. Within the Standard 
Model, the weak $B^0_s$--$\overline{B^0_s}$ mixing phase is very small, and 
studies of $B_s$ decays involve very rapid $B^0_s$--$\overline{B^0_s}$ 
oscillations due to the large mass difference $\Delta M_s\equiv 
M_{\rm H}^{(s)}-M_{\rm L}^{(s)}$ between the mass eigenstates 
$B_s^{\rm H}$ (``heavy'') and $B_s^{\rm L}$ (``light''). Future $B$-physics 
experiments performed at hadron machines should be in a position to resolve 
these oscillations. Interestingly, in contrast to the $B_d$ case, there may 
be a sizeable width difference $\Delta\Gamma_s\equiv\Gamma_{\rm H}^{(s)}-
\Gamma_{\rm L}^{(s)}$ between the mass eigenstates of the $B_s$ system 
\cite{DGamma}, which may allow studies of CP violation with ``untagged'' 
$B_s$ data samples, where one does not distinguish between initially, 
i.e.\ at time $t=0$, present $B_s^0$ or $\overline{B^0_s}$ mesons 
\cite{dunietz}. In such untagged rates, the rapid $B^0_s$--$\overline{B^0_s}$ 
oscillations cancel.

Some of the $B_s$ strategies proposed in the literature are theoretically 
clean, and use pure ``tree'' decays, for example $B_s\to D_s^\pm K^\mp$ 
\cite{Bsclean}. Since no flavour-changing neutral-current (FCNC) 
processes contribute to the corresponding decay amplitudes, it is quite
unlikely that they are significantly affected by new physics. Consequently, 
the preferred mechanism for physics beyond the Standard Model to manifest 
itself in the corresponding time-dependent decay rates is through 
contributions to $B^0_s$--$\overline{B^0_s}$ mixing. In contrast, the 
decay $B_s\to J/\psi\, K_{\rm S}$ discussed in this paper exhibits also
CP-violating effects that are due to the interference between ``tree'' and 
``penguin'', i.e.\ FCNC, processes. Therefore, new physics may well show 
up in the corresponding CP asymmetries, thereby affecting the extracted 
value of $\gamma$. A similar comment applies to the 
$B_{d (s)}\to D^{\,+}_{d(s)}\, D^{\,-}_{d(s)}$ strategy.

The outline of this paper is as follows: in Section~\ref{DAO}, the 
$B_{d(s)}\to J/\psi\, K_{\rm S}$ decay amplitudes are parametrized
in a completely general way within the framework of the Standard Model.
Moreover, expressions for the observables of the corresponding 
time-dependent decay rates are given. The strategy to determine $\gamma$
with the help of these observables is discussed in Section~\ref{gam-det},
whereas we turn to the analogous strategy using 
$B_{d (s)}\to D^{\,+}_{d(s)}\, D^{\,-}_{d(s)}$ decays in 
Section~\ref{BDDgam-det}. The main results are summarized in 
Section~\ref{sum}.

\begin{figure}
\begin{center}
\leavevmode
\epsfysize=5.5truecm 
\epsffile{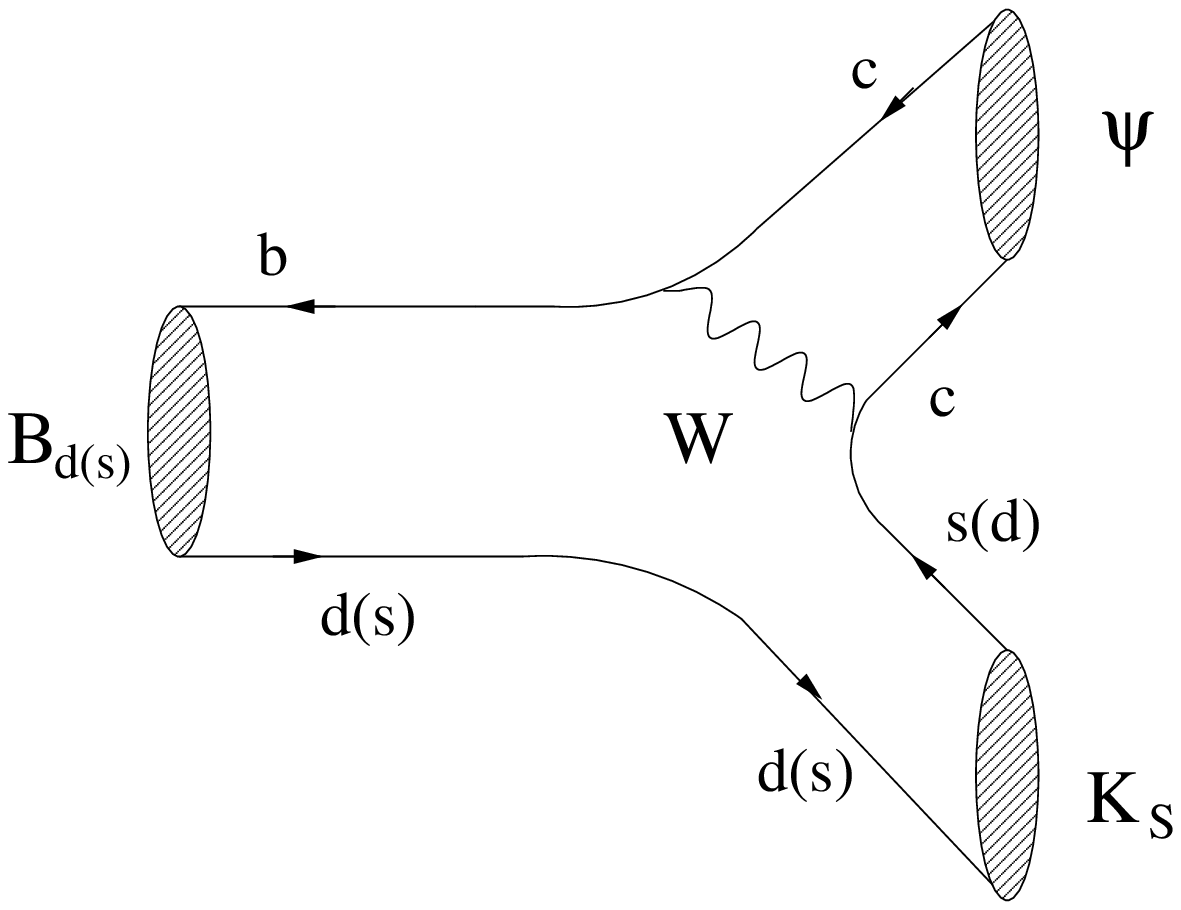} \hspace*{1truecm}
\epsfysize=5.5truecm 
\epsffile{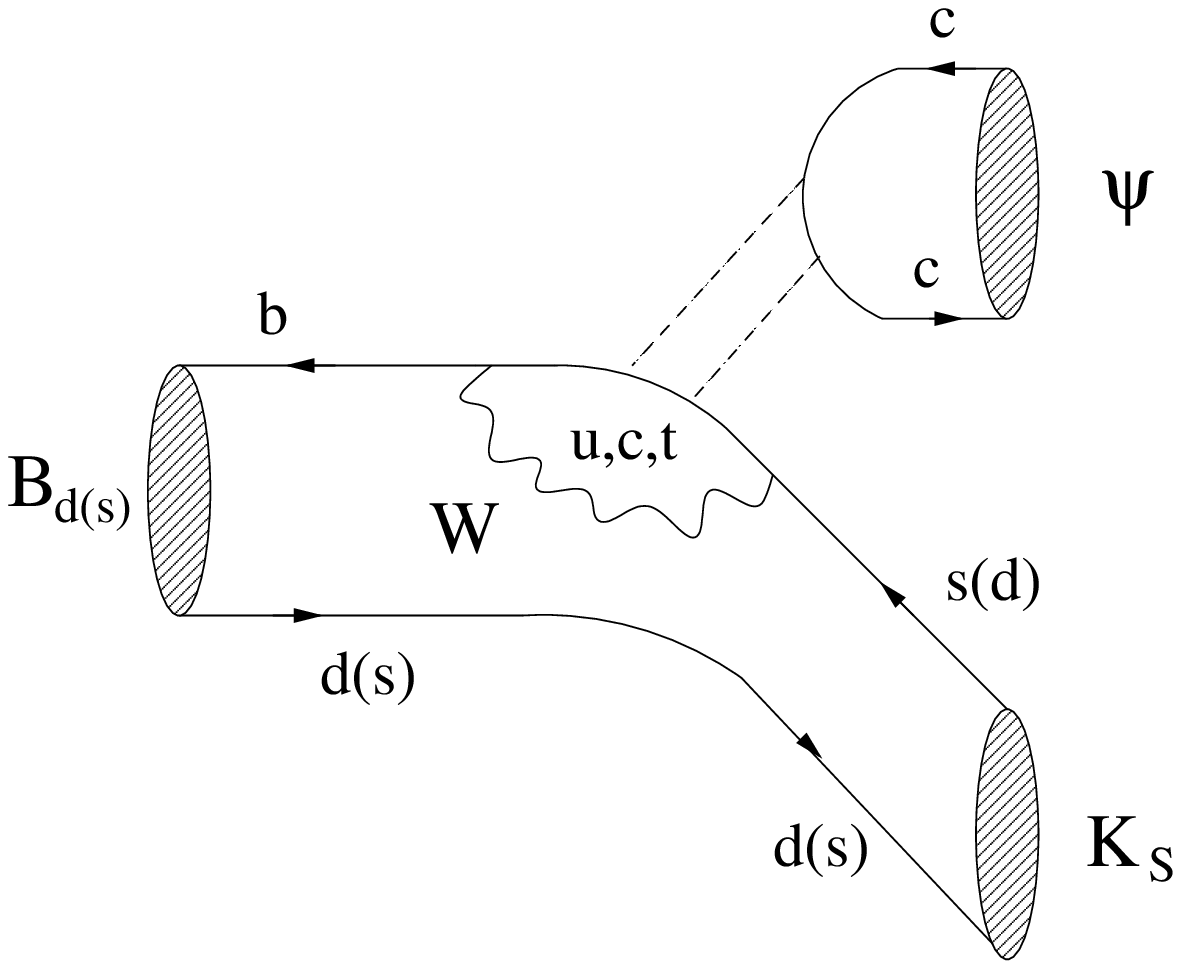}
\end{center}
\caption{Feynman diagrams contributing to $B_{d(s)}\to J/\psi\,K_{\rm S}$. 
The dashed lines in the penguin topology represent a colour-singlet 
exchange.}\label{fig:BdPsiKS}
\end{figure}

\boldmath
\section{The $B_{d(s)}\to J/\psi\, K_{\rm S}$ Observables}\label{DAO}
\unboldmath
The decays $B_{d(s)}^0\to J/\psi\, K_{\rm S}$ are transitions 
into a CP eigenstate with eigenvalue $-1$ and originate from 
$\bar b\to\bar cc\bar s (\bar d)$ quark-level decays. We have to 
deal both with current--current and with penguin contributions, as can be
seen in Fig.\ \ref{fig:BdPsiKS}. Let us turn to the mode
$B_d^0\to J/\psi\, K_{\rm S}$ first. Its transition amplitude can be
written as
\begin{equation}\label{Bd-ampl1}
A(B_d^0\to J/\psi\, K_{\rm S})=\lambda_c^{(s)}\left(A_{\rm cc}^{c'}+
A_{\rm pen}^{c'}\right)+\lambda_u^{(s)}A_{\rm pen}^{u'}
+\lambda_t^{(s)}A_{\rm pen}^{t'}\,,
\end{equation}
where $A_{\rm cc}^{c'}$ denotes the current--current contributions,
i.e.\ the ``tree'' processes in Fig.\ \ref{fig:BdPsiKS}, and the amplitudes 
$A_{\rm pen}^{q'}$ describe the contributions from penguin topologies with 
internal $q$ quarks ($q\in\{u,c,t\})$. These penguin amplitudes take into 
account both QCD and electroweak penguin contributions. The primes in 
(\ref{Bd-ampl1}) remind us that we are dealing with a $\bar b\to\bar s$
transition, and
\begin{equation}
\lambda_q^{(s)}\equiv V_{qs}V_{qb}^\ast
\end{equation}
are the usual CKM factors. Making use of the unitarity of the CKM matrix
and applying the Wolfenstein parametrization \cite{wolf}, generalized to
include non-leading terms in $\lambda$ \cite{blo}, we obtain 
\begin{equation}\label{Bd-ampl2}
A(B_d^0\to J/\psi\, K_{\rm S})=\left(1-\frac{\lambda^2}{2}\right){\cal A'}
\left[1+\left(\frac{\lambda^2}{1-\lambda^2}\right)a'e^{i\theta'}e^{i\gamma}
\right],
\end{equation}
where
\begin{equation}\label{Aap-def}
{\cal A'}\equiv\lambda^2A\left(A_{\rm cc}^{c'}+A_{\rm pen}^{ct'}\right),
\end{equation}
with $A_{\rm pen}^{ct'}\equiv A_{\rm pen}^{c'}-A_{\rm pen}^{t'}$, and
\begin{equation}\label{ap-def}
a'e^{i\theta'}\equiv R_b\left(1-\frac{\lambda^2}{2}\right)\left(
\frac{A_{\rm pen}^{ut'}}{A_{\rm cc}^{c'}+A_{\rm pen}^{ct'}}\right).
\end{equation}
The quantity $A_{\rm pen}^{ut'}$ is defined in analogy to $A_{\rm pen}^{ct'}$,
and the relevant CKM factors are given as follows:
\begin{equation}\label{CKM-exp}
\lambda\equiv|V_{us}|=0.22\,,\quad A\equiv\frac{1}{\lambda^2}
\left|V_{cb}\right|=0.81\pm0.06\,,\quad R_b\equiv\frac{1}{\lambda}
\left|\frac{V_{ub}}{V_{cb}}\right|=0.41\pm0.07\,.
\end{equation}

The decay $B_s^0\to J/\psi\, K_{\rm S}$ is related to $B_d^0\to J/\psi\, 
K_{\rm S}$ by interchanging all down and strange quarks, i.e.\ through
the so-called $U$-spin subgroup of the $SU(3)$ flavour symmetry of strong 
interactions. Using again the unitarity of the CKM matrix and a notation
similar to that in (\ref{Bd-ampl2}), we obtain
\begin{equation}\label{Bs-ampl}
A(B_s^0\to J/\psi\, K_{\rm S})=-\lambda\,{\cal A}\left[1-a\, e^{i\theta}
e^{i\gamma}\right],
\end{equation}
where
\begin{equation}
{\cal A}\equiv\lambda^2A\left(A_{\rm cc}^{c}+A_{\rm pen}^{ct}\right)
\end{equation}
and
\begin{equation}\label{a-def}
a\, e^{i\theta}\equiv R_b\left(1-\frac{\lambda^2}{2}\right)\left(
\frac{A_{\rm pen}^{ut}}{A_{\rm cc}^{c}+A_{\rm pen}^{ct}}\right)
\end{equation}
correspond to (\ref{Aap-def}) and (\ref{ap-def}), respectively. It should
be emphasized that (\ref{Bd-ampl2}) and (\ref{Bs-ampl}) are completely 
general parametrizations of the $B_{d(s)}^0\to J/\psi\, K_{\rm S}$ decay 
amplitudes within the Standard Model, relying only on the unitarity of 
the CKM matrix. In particular, these expressions also take into account 
final-state-interaction effects, which can be considered as long-distance 
penguin topologies with internal up- and charm-quark exchanges 
\cite{bfm,rf-BpiK}.

If we compare (\ref{Bd-ampl2}) and (\ref{Bs-ampl}) with each other, we
observe that the quantity $a' e^{i\theta'}$ is doubly Cabibbo-suppressed
in the $B_d^0\to J/\psi\, K_{\rm S}$ decay amplitude (\ref{Bd-ampl2}), 
whereas $a\, e^{i\theta}$ enters in the $B_s^0\to J/\psi\, K_{\rm S}$ 
amplitude (\ref{Bs-ampl}) in a Cabibbo-allowed way. This feature has important
implications for the CP-violating effects arising in the corresponding
time-dependent decay rates.

The time evolution for decays of initially, i.e.\ at time $t=0$, 
present neutral $B$ or $\overline{B}$ mesons into a final CP eigenstate 
$|f\rangle$, satisfying 
\begin{equation}
({\cal CP})|f\rangle=\eta\,|f\rangle, 
\end{equation}
is given as follows 
\cite{revs}:
\begin{equation}\label{At}
|A(t)|^2=\frac{|{\cal N}|^2}{2}\left[R_{\rm L}\,e^{-\Gamma_{\rm L}t}
+R_{\rm H}\,e^{-\Gamma_{\rm H}t}
+2\,e^{-\Gamma t}\left\{A_{\rm D}\cos(\Delta Mt) + A_{\rm M}\sin(\Delta Mt)
\right\}\right]
\end{equation}
\begin{equation}\label{Atbar}
|\overline{A}(t)|^2=\frac{|{\cal N}|^2}{2}\left[R_{\rm L}\,
e^{-\Gamma_{\rm L}t}+R_{\rm H}\,e^{-\Gamma_{\rm H}t}
-2\,e^{-\Gamma t}\left\{A_{\rm D}\cos(\Delta Mt) + A_{\rm M}\sin(\Delta Mt)
\right\}\right],
\end{equation}
where the $\Gamma_{\rm L,H}$ denote the decay widths of the $B$ mass 
eigenstates, $\Gamma\equiv(\Gamma_{\rm L}+\Gamma_{\rm H})/2$, and 
$\Delta M\equiv M_{\rm H}-M_{\rm L}>0$ is their mass difference. 
For the $B$ decays considered in this paper, the ``unevolved''
decay amplitudes take the form
\begin{eqnarray}
A&=&{\cal N}\left[1-b\, e^{i\rho} e^{+i\gamma}\right]\equiv {\cal N}z
\label{z-def}\\
\overline{A}&=&\eta\,{\cal N}\left[1-b\, e^{i\rho} e^{-i\gamma}\right]\equiv 
\eta\,{\cal N}\,\overline{z}\,,\label{zbar-def}
\end{eqnarray}
and we have
\begin{eqnarray}
\lefteqn{R_{\rm L}\equiv\frac{1}{2}\left[|z|^2+|\overline{z}|^2+2\,\eta\,
\Re\left(e^{-i\phi}z^\ast\,\overline{z}\right)\right]}\nonumber\\
&&=(1+\eta\cos\phi)-2\,b\cos\rho\left[\cos\gamma+\eta\cos(\phi+\gamma)
\right]+b^2\left[1+\eta\cos(\phi+2\,\gamma)\right]
\end{eqnarray}
\begin{eqnarray}
\lefteqn{R_{\rm H}\equiv\frac{1}{2}\left[|z|^2+|\overline{z}|^2-2\,\eta\,
\Re\left(e^{-i\phi}z^\ast\,\overline{z}\right)\right]}\nonumber\\
&&=(1-\eta\cos\phi)-2\,b\cos\rho\left[\cos\gamma-\eta\cos(\phi+\gamma)
\right]+b^2\left[1-\eta\cos(\phi+2\,\gamma)\right]
\end{eqnarray}
\vspace*{0.15truecm}
\begin{equation}
A_{\rm D}\equiv\frac{1}{2}\left(|z|^2-|\overline{z}|^2\right)=2\,b\,
\sin\rho\,\sin\gamma
\end{equation}
\vspace*{0.15truecm}
\begin{equation}
A_{\rm M}\equiv-\,\eta\,\Im\left(e^{-i\phi}z^\ast\,\overline{z}\right)=
\eta\left[\sin\phi-2\,b\,\cos\rho\,\sin(\phi+\gamma)+
b^2\sin(\phi+2\,\gamma)
\right].
\end{equation}
Here the phase $\phi$ denotes the $B$--$\overline{B}$ mixing phase:
\begin{equation}
\phi=\left\{
\begin{array}{cc}
2\beta&\mbox{$B_d$ system}\\
-2\delta\gamma&\mbox{$B_s$ system,}
\end{array}
\right.
\end{equation}
where $2\delta\gamma\approx0.03$ is tiny in the Standard Model 
because of a Cabibbo suppression of ${\cal O}(\lambda^2)$.
Note that the observables $R_{\rm L}$, $R_{\rm H}$, $A_{\rm D}$ and 
$A_{\rm M}$ satisfy the relation
\begin{equation}
A_{\rm D}^2+A_{\rm M}^2=R_{\rm L}R_{\rm H}.
\end{equation}

For the following considerations, it is useful to introduce the 
time-dependent CP asymmetry
\begin{equation}\label{acp-def}
a_{\rm CP}(t)\equiv\frac{|A(t)|^2-|\overline{A}(t)|^2}{|A(t)|^2+
|\overline{A}(t)|^2}=2\,e^{-\Gamma t}\left[\frac{{\cal A}_{\rm CP}^{\rm dir}
\cos(\Delta M t)+{\cal A}_{\rm CP}^{\rm mix}\sin(\Delta M t)}{
e^{-\Gamma_{\rm H}t}+e^{-\Gamma_{\rm L}t}+{\cal A}_{\rm \Delta\Gamma}\left(
e^{-\Gamma_{\rm H}t}-e^{-\Gamma_{\rm L}t}\right)} \right]
\end{equation}
with
\begin{eqnarray}
{\cal A}_{\rm CP}^{\rm dir}&\equiv&\frac{2A_{\rm D}}{R_{\rm H}+R_{\rm L}}=
\frac{2\,b\sin\rho\sin\gamma}{1-2\,b\cos\rho\cos\gamma+b^2}
\label{Adir-def}\\
{\cal A}_{\rm CP}^{\rm mix}&\equiv&\frac{2A_{\rm M}}{R_{\rm H}+R_{\rm L}}=
+\,\eta\left[\,\frac{\sin\phi-2\,b\,\cos\rho\,\sin(\phi+\gamma)+
b^2\sin(\phi+2\,\gamma)}{1-2\,b\cos\rho\cos\gamma+b^2}\,\right]
\label{Amix-def}\\
{\cal A}_{\Delta\Gamma}&\equiv&\frac{R_{\rm H}-R_{\rm L}}{R_{\rm H}+R_{\rm L}}=
-\,\eta\left[\frac{\cos\phi-2\,b\,\cos\rho\,\cos(\phi+\gamma)+
b^2\cos(\phi+2\,\gamma)}{1-2\,b\cos\rho\cos\gamma+b^2}\right],\label{AG-def}
\end{eqnarray}
and the observable
\begin{equation}\label{R-def}
R\equiv\frac{1}{2}\left(R_{\rm H}+R_{\rm L}\right)=1-
2\,b\cos\rho\cos\gamma+b^2.
\end{equation}
In the CP asymmetry (\ref{acp-def}), we have separated the ``direct'' from 
the ``mixing-induced'' CP-violating contributions. It is interesting to 
note that not only ${\cal A}_{\rm CP}^{\rm dir}$, but also $R$ does 
not depend on the $B$--$\overline{B}$ mixing phase $\phi$. The observables
${\cal A}_{\rm CP}^{\rm dir}$, ${\cal A}_{\rm CP}^{\rm mix}$ and
${\cal A}_{\Delta\Gamma}$ are not independent quantities, and satisfy the 
relation 
\begin{equation}\label{Obs-rel}
({\cal A}_{\rm CP}^{\rm dir})^2+({\cal A}_{\rm CP}^{\rm mix})^2
+({\cal A}_{\Delta\Gamma})^2=1.
\end{equation}

The formulae given above describe the time evolution of all kinds of 
neutral $B$ decays into a final CP eigenstate, where the ``unevolved'' 
decay amplitudes take the form specified in (\ref{z-def}) and 
(\ref{zbar-def}). Let us turn, in the following section, to the 
$B_{s(d)}\to J/\psi\, K_{\rm S}$ observables, which may provide an 
interesting strategy to determine $\gamma$.

\boldmath
\section{Extracting $\gamma$ from $B_{s(d)}\to J/\psi\, K_{\rm S}$ 
Decays}\label{gam-det}
\unboldmath
The observables introduced in (\ref{Adir-def})--(\ref{AG-def}) can be 
obtained directly from the time evolution of the decay rates corresponding
to (\ref{At}) and (\ref{Atbar}) and do not depend on the overall 
normalization $|{\cal N}|^2$. However, owing to (\ref{Obs-rel}), 
we have only two independent observables, depending on the three 
``unknowns'' $b$, $\rho$ and $\gamma$, and on the $B$--$\overline{B}$
mixing phase $\phi$. Consequently, in order to determine these ``unknowns'',
we need an additional observable, which is provided by $R$. Unfortunately, 
the time-dependent decay rates fix only the quantity
\begin{equation}\label{RN-def}
\langle\Gamma\rangle\equiv\mbox{PhSp}\times|{\cal N}|^2\times R=
\mbox{PhSp}\times|{\cal N}|^2\times\frac{1}{2}\,(R_{\rm H}+R_{\rm L})
\end{equation}
through
\begin{equation}\label{untag-rate}
\Gamma(B(t)\to f)+\Gamma(\overline{B}(t)\to f)=\mbox{PhSp}\times|{\cal N}|^2
\times\left[R_{\rm H}e^{-\Gamma_{\rm H}t}+R_{\rm L}e^{-\Gamma_{\rm L}t}\right],
\end{equation}
where ``PhSp'' denotes an appropriate, straightforwardly calculable 
phase-space factor. Consequently, the overall normalization 
$|{\cal N}|^2$ is required in order to determine $R$. In the case of the 
decay $B_s\to J/\psi\, K_{\rm S}$, this normalization can be fixed through 
the CP-averaged $B_d\to J/\psi\, K_{\rm S}$ rate with the help of the 
$U$-spin symmetry. 

In the case of $B_d\to J/\psi\, K_{\rm S}$, we have
\begin{equation}
{\cal N}=\left(1-\frac{\lambda^2}{2}\right){\cal A}',\quad
b=\epsilon\,a',\quad\rho=\theta'+180^\circ,
\quad\mbox{with}\quad\epsilon\equiv\frac{\lambda^2}{1-\lambda^2}\,,
\end{equation}
whereas we have in the $B_s\to J/\psi\, K_{\rm S}$ case
\begin{equation}
{\cal N}=-\lambda\,{\cal A},\quad b=a,\quad\rho=\theta.
\end{equation}
Consequently, we obtain
\begin{equation}\label{H-def}
H\equiv\frac{1}{\epsilon}\left(\frac{|{\cal A}'|}{|{\cal A}|}\right)^2
\left[\frac{M_{B_d}\Phi(M_{J/\psi}/M_{B_d},
M_K/M_{B_d})}{M_{B_s}\Phi(M_{J/\psi}/M_{B_s},M_K/M_{B_s})}\right]^3
\frac{\langle\Gamma\rangle}{\langle\Gamma'\rangle}=
\frac{1-2\,a\,\cos\theta\cos\gamma+a^2}{1+
2\,\epsilon\,a'\cos\theta'\cos\gamma+\epsilon^2\,a'^2}\,,
\end{equation}
where 
\begin{equation}\label{PhSp}
\Phi(x,y)=\sqrt{\left[1-(x+y)^2\right]\left[1-(x-y)^2\right]}
\end{equation}
is the usual two-body phase-space function, and 
$\langle\Gamma\rangle\equiv \langle\Gamma(B_s\to J/\psi\, K_{\rm S})\rangle$ 
and $\langle\Gamma'\rangle\equiv \langle\Gamma(B_d\to J/\psi\, K_{\rm S})
\rangle$ can be determined from the ``untagged'' 
$B_{s(d)}\to J/\psi\, K_{\rm S}$ rates with the help of (\ref{RN-def}) and 
(\ref{untag-rate}). Since the $U$-spin flavour symmetry of strong 
interactions implies
\begin{equation}\label{SU3-1}
|{\cal A}'|=|{\cal A}|
\end{equation}
and
\begin{equation}\label{SU3-2}
a'=a,\quad \theta'=\theta,
\end{equation}
we can determine $a$, $\theta$ and $\gamma$ as a function of the
$B_s^0$--$\overline{B_s^0}$ mixing phase by combining $H$ with 
${\cal A}_{\rm CP}^{\rm dir}\equiv
{\cal A}_{\rm CP}^{\rm dir}(B_s\to J/\psi\, K_{\rm S})$ and
${\cal A}_{\rm CP}^{\rm mix}\equiv
{\cal A}_{\rm CP}^{\rm mix}(B_s\to J/\psi\, K_{\rm S})$ or
${\cal A}_{\Delta\Gamma}\equiv {\cal A}_{\Delta\Gamma}
(B_s\to J/\psi\, K_{\rm S})$. In contrast to certain isospin relations,
electroweak penguins do not lead to any problems in these $U$-spin 
relations. As we have already noted, 
the $B^0_s$--$\overline{B^0_s}$ mixing phase $\phi=-2\delta\gamma$ is 
expected to be negligibly small in the Standard Model. It can be
probed with the help of the decay $B_s\to J/\psi\,\phi$ (see, for example, 
\cite{ddf1}). Large CP-violating effects in this decay would signal that 
$2\delta\gamma$ is not tiny, and would indicate new-physics contributions 
to $B^0_s$--$\overline{B^0_s}$ mixing. Strictly speaking, in the case
of $B_s\to J/\psi\, K_{\rm S}$, we have $\phi=-2\delta\gamma-\phi_K$, where
$\phi_K$ is related to the $K^0$--$\overline{K^0}$ mixing phase and is
negligibly small in the Standard Model. On the other hand, we have
$\phi=2\beta+\phi_K$ in the case of $B_d\to J/\psi\, K_{\rm S}$. Since
the value of the CP-violating parameter $\varepsilon_K$ of the 
neutral kaon system is small, $\phi_K$ can only be affected by very 
contrived models of new physics \cite{nirsil}.

An important by-product of the strategy described above is that 
the quantities $a'$ and $\theta'$ allow us to take into account the penguin 
contributions in the determination of $\beta$ from $B_d\to J/\psi\, 
K_{\rm S}$, which are presumably very small because of the Cabibbo suppression 
of $\lambda^2/(1-\lambda^2)$ in (\ref{Bd-ampl2}). Moreover, using 
(\ref{SU3-2}), we obtain an interesting relation between the direct CP 
asymmetries arising in the modes $B_d\to J/\psi\, K_{\rm S}$ and 
$B_s\to J/\psi\, K_{\rm S}$ and their CP-averaged rates:
\begin{equation}
\frac{{\cal A}_{\rm CP}^{\rm dir}(B_d\to J/\psi\, 
K_{\rm S})}{{\cal A}_{\rm CP}^{\rm dir}(B_s\to J/\psi\, K_{\rm S})}=
-\,\epsilon\,H=-\left(\frac{|{\cal A}'|}{|{\cal A}|}\right)^2
\left[\frac{M_{B_d}\Phi(M_{J/\psi}/M_{B_d},
M_K/M_{B_d})}{M_{B_s}\Phi(M_{J/\psi}/M_{B_s},M_K/M_{B_s})}\right]^3
\frac{\langle\Gamma\rangle}{\langle\Gamma'\rangle}.
\end{equation}
An analogous relation holds also between the $B^\pm\to\pi^\pm K$ and
$B^\pm\to K^\pm K$ CP-violating asymmetries \cite{bfm,rf-BpiK}. At 
``second-generation'' $B$-physics experiments at hadron machines, for 
instance at LHCb, the sensitivity may be good enough to resolve a direct CP 
asymmetry in $B_d\to J/\psi\, K_{\rm S}$. In view of the impressive accuracy 
that can be achieved in the era of such experiments, it is also an important
issue to think about the theoretical accuracy of the determination of $\beta$
from $B_d\to J/\psi\, K_{\rm S}$. The approach discussed above allows us
to control these -- presumably very small -- hadronic uncertainties with 
the help of $B_s\to J/\psi\, K_{\rm S}$.

Interestingly, the strategy to extract $\gamma$ from $B_{s(d)}\to J/\psi\, 
K_{\rm S}$ decays does not require a non-trivial CP-conserving strong phase 
$\theta$. However, its experimental feasibility depends strongly on the 
value of the quantity $a$ introduced in (\ref{a-def}). It is very difficult 
to estimate $a$ theoretically. In contrast to the ``usual'' QCD penguin 
topologies, the QCD penguins contributing to $B_{s(d)}\to J/\psi\, K_{\rm S}$
require a colour-singlet exchange, as indicated in Fig.~\ref{fig:BdPsiKS}
through the dashed lines, and are ``Zweig-suppressed''. Such a comment does 
not apply to the electroweak penguins, which contribute in ``colour-allowed'' 
form. The current--current amplitude $A_{\rm cc}^c$ is due to 
``colour-suppressed'' topologies, and the ratio 
$A_{\rm pen}^{ut}/(A_{\rm cc}^{c}+A_{\rm pen}^{ct})$, which governs $a$, 
may be sizeable. It is interesting to note that the measured branching 
ratio $\mbox{BR}(B^0_d\to J/\psi\, K^0)=
2\,\mbox{BR}(B_d^0\to J/\psi\, K_{\rm S})=(8.9\pm1.2)\times10^{-4}$ \cite{PDG} 
probes only the combination ${\cal A}'\propto\left(A_{\rm cc}^{c'}+
A_{\rm pen}^{ct'}\right)$ of current--current and penguin amplitudes, 
and obviously does not allow us to separate these contributions. It would
be very important to have a better theoretical understanding of the
quantity $a\,e^{i\theta}$. However, such analyses are far beyond the 
scope of this paper, and are left for further studies. If we use 
\begin{equation}
\frac{\mbox{BR}(B_s\to J/\psi\, K_{\rm S})}{\mbox{BR}(B_d\to J/\psi\, 
K_{\rm S})}=\epsilon\, H\left(\frac{|{\cal A}|}{|{\cal A}'|}\right)^2
\left[\frac{M_{B_s}\Phi(M_{J/\psi}/M_{B_s},
M_K/M_{B_s})}{M_{B_d}\Phi(M_{J/\psi}/M_{B_d},M_K/M_{B_d})}\right]^3
\frac{\tau_{B_s}}{\tau_{B_d}}
\end{equation}
and (\ref{SU3-1}), we expect a $B_s\to J/\psi\, K_{\rm S}$ branching ratio 
at the level of $2\times10^{-5}$.

The general expressions for the observables (\ref{Adir-def})--(\ref{AG-def})
and (\ref{H-def}) are quite complicated. However, they simplify considerably, 
if we keep only the terms linear in $a$. Within this approximation, we obtain 
the simple result 
\begin{equation}\label{gam-approx}
\tan\gamma\approx\frac{\sin\phi-\eta{\cal A}_{\rm CP}^{\rm mix}}{(1-H)
\cos\phi}=\left.-\left(\frac{\eta{\cal A}_{\rm CP}^{\rm mix}}{1-H}\right)
\right|_{\phi=0},
\end{equation}
allowing us to determine $\gamma$ from the CP-averaged 
$B_{s(d)}\to J/\psi\, K_{\rm S}$ rates and the mixing-induced CP asymmetry 
arising in $B_s\to J/\psi\, K_{\rm S}$.

In the general case, where no approximations are made, there is also a 
``transparent'' strategy to determine $\gamma$. The point is that the 
CP-violating asymmetries ${\cal A}_{\rm CP}^{\rm dir}$ and 
${\cal A}_{\rm CP}^{\rm mix}$ allow us to fix contours in the $\gamma$--$a$ 
plane, which are described by
\begin{equation}\label{contours1}
a=\sqrt{\frac{1}{k}\left[l\pm\sqrt{l^2-h\,k}\right]},
\end{equation}
where
\begin{eqnarray}
h&=&u^2+D\,(1-u\cos\gamma)^2\\
k&=&v^2+D\,(1-v\cos\gamma)^2\\
l&=&2-u\,v-D\,(1-u\cos\gamma)(1-v\cos\gamma)
\end{eqnarray}
with 
\begin{eqnarray}
u&=&\frac{(\eta{\cal A}_{\rm CP}^{\rm mix})-\sin\phi}{
(\eta{\cal A}_{\rm CP}^{\rm mix})\cos\gamma-\sin(\phi+\gamma)}\label{u-def}\\
v&=&\frac{(\eta{\cal A}_{\rm CP}^{\rm mix})-\sin(\phi+2\,\gamma)}{(\eta
{\cal A}_{\rm CP}^{\rm mix})\cos\gamma-\sin(\phi+\gamma)}\label{v-def}
\end{eqnarray}
and
\begin{equation}
D=\left(\frac{{\cal A}_{\rm CP}^{\rm dir}}{\sin\gamma}\right)^2.
\end{equation}
It should be emphasized that these contours are {\it theoretically clean}.
It is also possible to combine the direct and mixing-induced
CP asymmetries arising in $B_d\to\pi^+\pi^-$ in an analogous way \cite{fm1}, 
allowing us to fix certain contours as well \cite{charles}.

\begin{figure}
\centerline{\rotate[r]{
\epsfysize=11.1truecm
{\epsffile{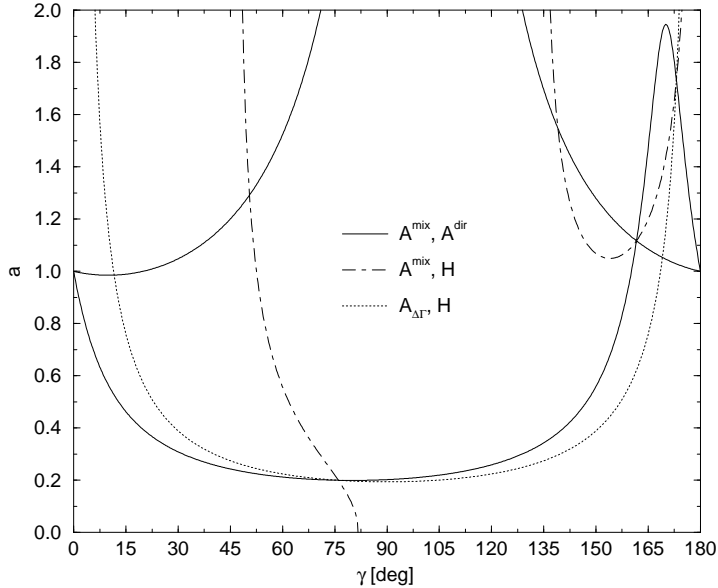}}}}
\caption{The contours in the $\gamma$--$a$ plane fixed through the 
$B_{s(d)}\to J/\psi\, K_{\rm S}$ observables for a specific example
discussed in the text.}\label{fig:Bscont}
\end{figure}

So far, we have not yet used the observable $H$. Combining it with 
${\cal A}_{\rm CP}^{\rm mix}$, we can fix another contour in the
$\gamma$--$a$ plane:
\begin{equation}\label{contours2}
a=\sqrt{\frac{H-1+u\,(1+\epsilon H)\cos\gamma}{1-v\,(1+\epsilon H)\cos\gamma
-\epsilon^2 H}}\,.
\end{equation}
If we use ${\cal A}_{\Delta\Gamma}$ instead of ${\cal A}_{\rm CP}^{\rm mix}$,
we obtain the same expression for $a$ as given in (\ref{contours2}), where 
$u$ and $v$ specified in (\ref{u-def}) and (\ref{v-def}) are replaced by
\begin{eqnarray}
u&\to&\frac{(\eta{\cal A}_{\Delta\Gamma})+\cos\phi}{
(\eta{\cal A}_{\Delta\Gamma})\cos\gamma+\cos(\phi+\gamma)}\\
v&\to&\frac{(\eta{\cal A}_{\Delta\Gamma})+\cos(\phi+2\,\gamma)}{(\eta
{\cal A}_{\Delta\Gamma})\cos\gamma+\cos(\phi+\gamma)}.
\end{eqnarray}

The intersection of the contours described by (\ref{contours1}) and
(\ref{contours2}) fixes both $a$ and $\gamma$. Let us illustrate this
approach in a quantitative way by considering a simple example. Assuming
a negligible $B^0_s$--$\overline{B^0_s}$ mixing phase, i.e.\ $\phi=0$, 
and $\gamma=76^\circ$, which lies within the range allowed at present for 
this angle, implied by the usual indirect fits of the unitarity triangle,
as well as $a=a'=0.2$ and $\theta=\theta'=30^\circ$, we obtain the 
$B_s\to J/\psi\, K_{\rm S}$ observables ${\cal A}_{\rm CP}^{\rm dir}=0.20$, 
${\cal A}_{\rm CP}^{\rm mix}=0.33$, ${\cal A}_{\Delta\Gamma}=0.92$ and 
$H=0.95$. The corresponding contours in the $\gamma$--$a$ plane are shown 
in Fig.~\ref{fig:Bscont}, where the solid lines are obtained with the 
help of (\ref{contours1}), and the dot-dashed lines
correspond to (\ref{contours2}). Interestingly, in the case of the contours
shown in Fig.~\ref{fig:Bscont}, we would not have to deal with ``physical'' 
discrete ambiguities for $\gamma$, since values of $a$ larger than 1 would
simply appear unrealistic. If it should become possible to measure 
${\cal A}_{\Delta\Gamma}$ with the help of the widths difference
$\Delta\Gamma_s$, the dotted line could be fixed. In this example, the
approximate expression (\ref{gam-approx}) yields $\gamma\approx82^\circ$,
which deviates from the ``true'' value of $\gamma=76^\circ$ by only 8\%.
It is also interesting to note that we have 
${\cal A}_{\rm CP}^{\rm dir}(B_d\to J/\psi\, K_{\rm S})=-0.98\%$ in our 
example.

\begin{figure}
\begin{center}
\leavevmode
\epsfysize=5truecm 
\epsffile{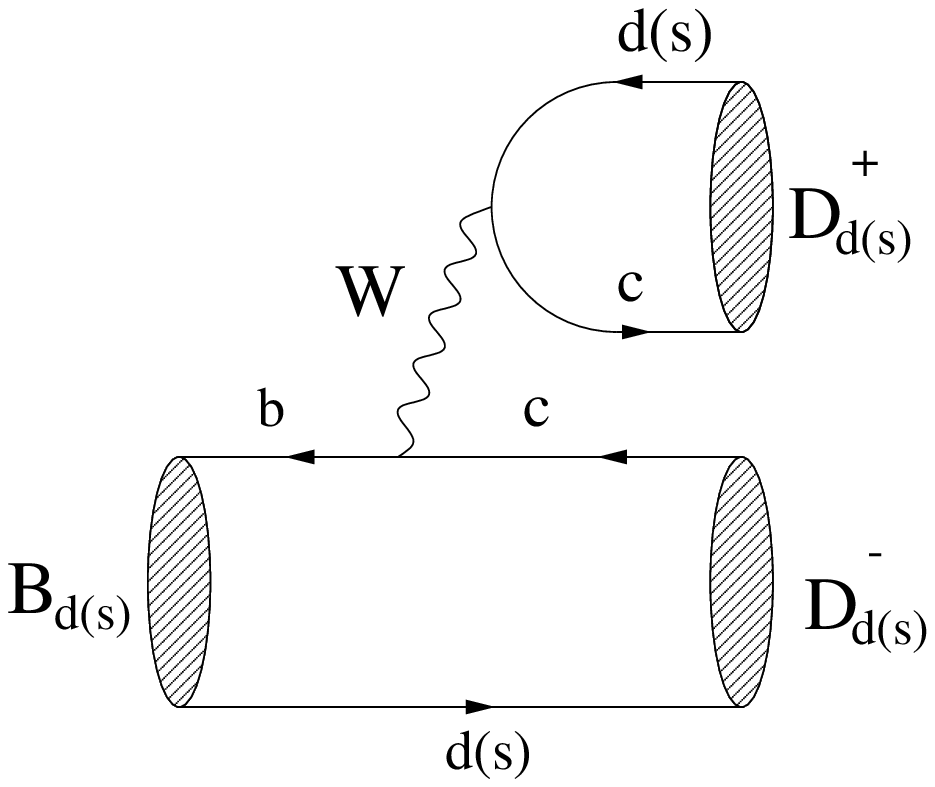} \hspace*{2truecm}
\epsfysize=5.5truecm 
\epsffile{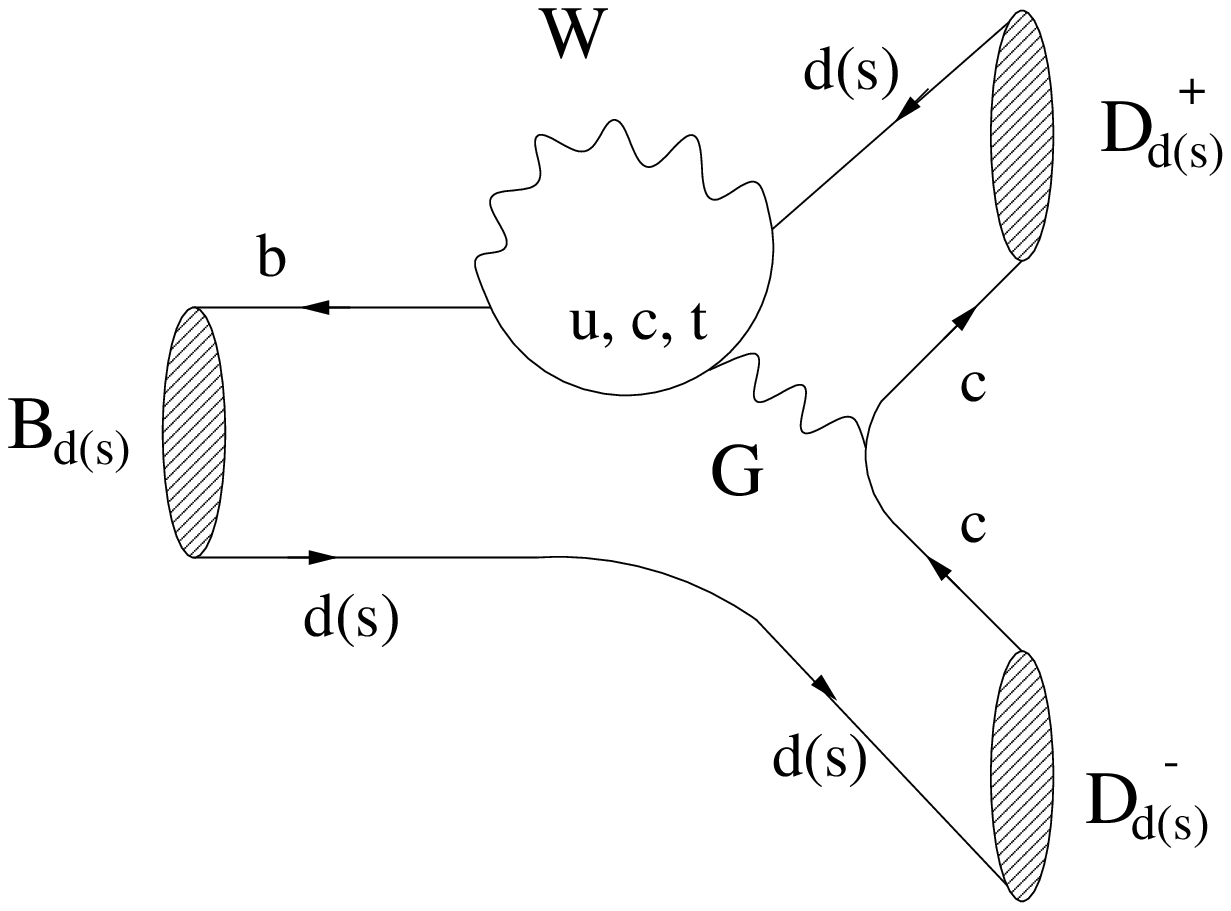}
\end{center}
\caption{Feynman diagrams contributing to 
$B_{d (s)}\to D^{\,+}_{d(s)}\, D^{\,-}_{d(s)}$.}\label{fig:BDD}
\end{figure}

Before turning to the $B_{d (s)}\to D^{\,+}_{d(s)}\, D^{\,-}_{d(s)}$ decays
in the next section, let us say a few words on the $SU(3)$-breaking
corrections. Whereas the contours in the $\gamma$--$a$ plane related
to (\ref{contours1}), i.e.\ the solid curves in Fig.\ \ref{fig:Bscont}, 
are {\it theoretically clean}, those described by (\ref{contours2}), i.e.\ 
the dot-dashed and dotted lines in Fig.\ \ref{fig:Bscont}, are affected 
by $U$-spin-breaking corrections. Because of the small parameter 
$\epsilon=0.05$ in (\ref{H-def}), these contours are essentially unaffected 
by possible corrections to (\ref{SU3-2}), and rely predominantly on the 
$U$-spin relation $|{\cal A'}|=|{\cal A}|$. In the ``factorization'' 
approximation, we have
\begin{equation}\label{SU3-break}
\left.\frac{|{\cal A'}|}{|{\cal A}|}\right|_{\rm fact}=\,
\frac{F_{B_d^0K^0}(M_{J/\psi}^2;1^-)}{F_{B_s^0\overline{K^0}}
(M_{J/\psi}^2;1^-)}\,,
\end{equation}
where the form factors $F_{B_d^0K^0}(M_{J/\psi}^2;1^-)$ and 
$F_{B_s^0\overline{K^0}}(M_{J/\psi}^2;1^-)$ parametrize the 
quark-current matrix elements 
$\langle K^0|(\bar b s)_{\rm V-A}|B^0_d\rangle$ and 
$\langle\overline{K^0}|(\bar b d)_{\rm V-A}|B^0_s\rangle$, respectively 
\cite{BSW}. We are not aware of quantitative
studies of (\ref{SU3-break}), which could be performed, for instance, with 
the help of sum rule or lattice techniques. In the light-cone sum-rule
approach, sizeable $SU(3)$-breaking effects were found in the case of the
$B_{d,s}\to K^\ast$ form factors \cite{BB}. It should be emphasized that 
also non-factorizable corrections, which are not included in 
(\ref{SU3-break}), may play an important role. We are optimistic that we 
will have a better picture of $SU(3)$ breaking by the time the 
$B_s\to J/\psi\, K_{\rm S}$ measurements can be performed in practice.

\boldmath
\section{Extracting $\gamma$ from $B_{d (s)}\to D^{\,+}_{d(s)}\, 
D^{\,-}_{d(s)}$ Decays}\label{BDDgam-det}
\unboldmath
The decays $B_{d (s)}^0\to D^{\,+}_{d(s)}\, D^{\,-}_{d(s)}$ are 
transitions into a CP eigenstate with eigenvalue $+1$ and originate from 
$\bar b\to\bar cc\bar d (\bar s)$ quark-level decays. We have to 
deal both with current--current and with penguin contributions, as can be
seen in Fig.\ \ref{fig:BDD}. In analogy to (\ref{Bd-ampl2}) and 
(\ref{Bs-ampl}), the corresponding transition amplitudes can be written as
\begin{eqnarray}
A(B_s^0\to D_s^+D_s^-)&=&\left(1-\frac{\lambda^2}{2}\right)\tilde{\cal A}'
\left[1+\left(\frac{\lambda^2}{1-\lambda^2}\right)\tilde a'e^{i\tilde
\theta'}e^{i\gamma}\right]\label{BDDs-ampl}\\
A(B_d^0\to D_d^+D_d^-)&=&-\lambda\,\tilde{\cal A}\left[1-\tilde a\, 
e^{i\tilde\theta}e^{i\gamma}\right],\label{BDDd-ampl}
\end{eqnarray}
where the quantities $\tilde{\cal A}$, $\tilde{\cal A}'$ and 
$\tilde a\,e^{i\tilde\theta}$, $\tilde a'\,e^{i\tilde\theta'}$ take the
same form as in the $B_{s (d)}\to J/\psi\, K_{\rm S}$ case. In contrast to
$B_{s(d)}\to J/\psi\, K_{\rm S}$, there are ``colour-allowed''
current--current contributions to $B_{d (s)}\to D^{\,+}_{d(s)}\, 
D^{\,-}_{d(s)}$, as well as contributions from ``exchange'' topologies, 
and the QCD penguins do not require a colour-singlet exchange, i.e.\
are not ``Zweig-suppressed''. 

Usually, $B_d\to D_d^+D_d^-$ decays appear in the literature as a tool
to probe $\beta$ \cite{revs}. In fact, if penguins played a negligible
role in these modes, $\beta$ could be determined from the corresponding
mixing-induced CP-violating effects. However, the penguin topologies,
which contain also important contributions from final-state-interaction
effects, may well be sizeable, although it is very difficult to calculate 
them in a reliable way. The strategy proposed here makes use of these penguin 
topologies, allowing us to determine $\gamma$, if the overall 
$B_d\to D_d^+D_d^-$ normalization is fixed through the CP-averaged, i.e.\ 
the ``untagged'' $B_s\to D_s^+D_s^-$ rate, and if the 
$B^0_d$--$\overline{B^0_d}$ mixing phase $2\beta$ is determined with the 
help of $B_d\to J/\psi\, K_{\rm S}$. 
It should be emphasized that no $\Delta M_st$ oscillations have to be 
resolved to measure the untagged $B_s\to D_s^+D_s^-$ rate. Since 
the phase structures of the $B_d^0\to D_d^+D_d^-$ and $B_s^0\to D_s^+D_s^-$ 
decay amplitudes are completely analogous to those of $B_s^0\to J/\psi\, 
K_{\rm S}$ and $B_d^0\to J/\psi\, K_{\rm S}$, respectively, the formalism 
developed in the previous section can be applied by performing straightforward
replacements of variables. Taking into account phase-space effects,
we have
\begin{equation}
\tilde H=\frac{1}{\epsilon}
\left(\frac{|\tilde{\cal A'}|}{|\tilde{\cal A}|}\right)^2
\left[\frac{M_{B_d}}{M_{B_s}}\frac{\Phi(M_{D_s}/M_{B_s},
M_{D_s}/M_{B_s})}{\Phi(M_{D_d}/M_{B_d},M_{D_d}/M_{B_d})}\right]
\frac{\langle\tilde\Gamma\rangle}{\langle\tilde\Gamma'\rangle}\,,
\end{equation}
where the CP-averaged rates $\langle\tilde\Gamma\rangle\equiv\langle\Gamma
(B_d\to D_d^+D_d^-)\rangle$ and $\langle\tilde\Gamma'\rangle\equiv\langle
\Gamma(B_s\to D_s^+D_s^-)\rangle$ can be determined with the help of 
(\ref{RN-def}) and (\ref{untag-rate}), and the function $\Phi(x,y)$ is as
given in (\ref{PhSp}).

\begin{figure}
\centerline{\rotate[r]{
\epsfysize=11.05truecm
{\epsffile{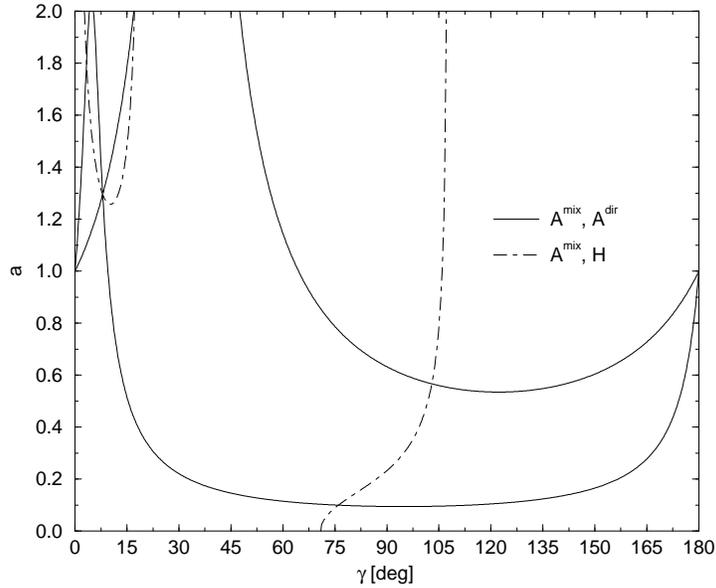}}}}
\caption{The contours in the $\gamma$--$\tilde a$ plane fixed through the 
$B_{d (s)}\to D^{\,+}_{d(s)}\, D^{\,-}_{d(s)}$ observables for a specific 
example discussed in the text ($2\beta=53^\circ$).}\label{fig:Bdcontours1}
\end{figure}

\begin{figure}
\centerline{\rotate[r]{
\epsfysize=11.05truecm
{\epsffile{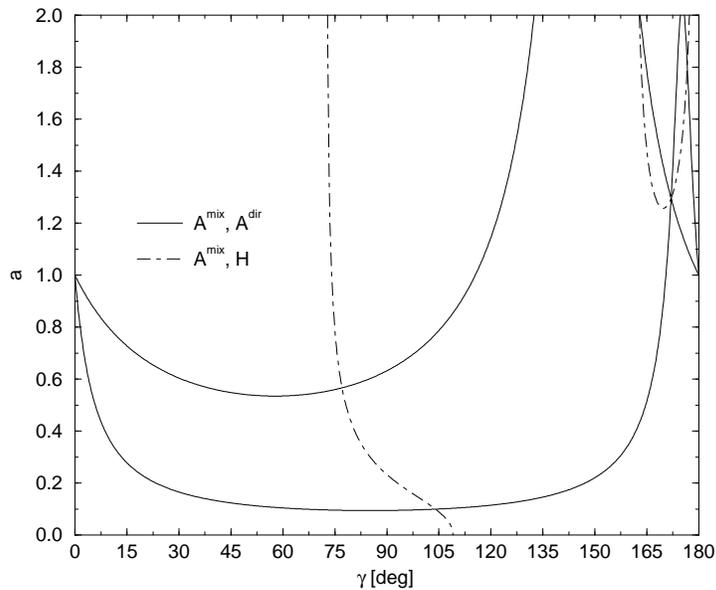}}}}
\caption{The contours in the $\gamma$--$\tilde a$ plane fixed through the 
$B_{d (s)}\to D^{\,+}_{d(s)}\, D^{\,-}_{d(s)}$ observables for a specific 
example discussed in the text 
($2\beta=180^\circ-53^\circ$).}\label{fig:Bdcontours2}
\end{figure}

Let us illustrate the strategy to determine $\gamma$, again by considering 
a simple example. Assuming $\tilde a=\tilde a'=0.1$, $\tilde\theta=
\tilde\theta'=210^\circ$, $\gamma=76^\circ$ and a $B^0_d$--$\overline{B^0_d}$ 
mixing phase of $\phi=2\beta=53^\circ$, we obtain the $B_d\to D_d^+D_d^-$ 
observables $\tilde{\cal A}_{\rm CP}^{\rm dir}=-0.092$, 
$\tilde{\cal A}_{\rm CP}^{\rm mix}=0.88$ and $\tilde H=1.05$. In this case, 
studies of CP violation in $B_d\to J/\psi\, K_{\rm S}$ would yield 
$\sin(2\beta)=0.8$, which is the central value of the most recent CDF 
analysis \cite{exp}, implying $2\beta=53^\circ$ or 
$2\beta=180^\circ-53^\circ=127^\circ$. 
Here we have assumed that $\beta\in[0^\circ,180^\circ]$, as implied by
the measured value of $\varepsilon_K$. A similar comment applies to the 
range for $\gamma$. The former solution for $2\beta$ would lead to the 
contours in the $\gamma$--$\tilde a$ plane shown in 
Fig.~\ref{fig:Bdcontours1}. The contours corresponding to $2\beta=127^\circ$ 
are shown in Fig.~\ref{fig:Bdcontours2}. Since values of 
$\tilde a={\cal O}(1)$ appear unrealistic, we would obtain the two 
``physical'' solutions of $76^\circ$ and $104^\circ$ for $\gamma$, 
which are due to the twofold ambiguity of $2\beta$. There are several 
strategies to resolve this discrete ambiguity in the extraction of 
$\beta$ \cite{ambig}, which should be feasible in the era of 
``second-generation'' $B$-physics experiments.

As in the $B_{s (d)}\to J/\psi\, K_{\rm S}$ case, only the contours 
involving the observable $\tilde H$ are affected by $SU(3)$-breaking 
corrections. Because of the small parameter $\epsilon$, they are essentially
due to the $U$-spin-breaking corrections to $|\tilde{\cal A}'|=
|\tilde{\cal A}|$. Within the ``factorization'' approximation, we have
\begin{equation}\label{SU3-breakBDD}
\left.\frac{|\tilde{\cal A'}|}{|\tilde{\cal A}|}\right|_{\rm fact}\approx\,
\frac{(M_{B_s}-M_{D_s})\,\sqrt{M_{B_s}M_{D_s}}\,(w_s+1)}{(M_{B_d}-M_{D_d})
\,\sqrt{M_{B_d}M_{D_d}}\,(w_d+1)}\frac{f_{D_s}\,\xi_s(w_s)}{f_{D_d}\,
\xi_d(w_d)}\,,
\end{equation}
where the restrictions form the heavy-quark effective theory for the 
$B_q\to D_q$ form factors have been taken into account by introducing 
appropriate Isgur--Wise functions $\xi_q(w_q)$ with $w_q=M_{B_q}/(2M_{D_q})$ 
\cite{neu-ste}. Studies of the light-quark dependence of the Isgur--Wise
function were performed within heavy-meson chiral perturbation theory, 
indicating an enhancement of $\xi_s/\xi_d$ at the level of $5\%$ 
\cite{HMChiPT1}. Applying the same formalism to $f_{D_s}/f_D$ gives values
at the 1.2 level \cite{HMChiPT2}, which is of the same order of magnitude as 
the results of recent lattice calculations \cite{lat1}. Further studies are 
needed to get a better picture of the $SU(3)$-breaking corrections 
to the ratio $|\tilde{\cal A'}|/|\tilde{\cal A}|$. Since ``factorization'' 
may work reasonably well for $B_q\to D_q^+D_q^-$, the leading corrections 
are expected to be due to (\ref{SU3-breakBDD}).

The experimental feasibility of the strategy to extract $\gamma$ from
$B_{d (s)}\to D^{\,+}_{d(s)}\, D^{\,-}_{d(s)}$ decays depends strongly
on the size of the penguin parameter $\tilde a$, which is difficult to 
predict theoretically. The branching ratio for $B_d^0\to D_d^+D_d^-$ is 
expected at the $4\times10^{-4}$ level \cite{neu-ste}; the one for 
$B_s^0\to D_s^+D_s^-$ is enhanced by $1/\epsilon=20$, and is correspondingly 
expected at the $8\times10^{-3}$ level. Already at the asymmetric 
$e^+e^-$ $B$-factories starting very soon, it should be possible to perform 
time-dependent measurements of the decay $B_d\to D_d^+D_d^-$, whereas 
$B_s\to D_s^+D_s^-$ -- and its ``untagged'' rate -- may be accessible at 
CDF or HERA--B.  However, unless the penguin effects in $B_d\to D_d^+D_d^-$ 
are very large, the approach to determine $\gamma$ discussed in this section 
appears to be particularly interesting for ``second-generation'' experiments, 
such as LHCb. The $e^+e^-$ $B$-factory experiments should nevertheless have 
a very careful look at the decay $B_d\to D_d^+D_d^-$, and those at hadron 
machines should study its $U$-spin counterpart $B_s\to D_s^+D_s^-$.

\section{Summary}\label{sum}
The observables of the time-dependent $B_s\to J/\psi\, K_{\rm S}$ rate, in 
combination with the CP-averaged $B_d\to J/\psi\, K_{\rm S}$ rate, provide 
an interesting strategy to determine the angle $\gamma$ of the unitarity 
triangle. This approach is not affected by any final-state-interaction 
effects, and its theoretical accuracy is only limited by the $U$-spin 
flavour symmetry of strong interactions. As a by-product, it allows us to 
take into account the penguin effects in the determination of $\beta$ from 
$B_d\to J/\psi\, K_{\rm S}$, which are presumably very small. An analogous 
strategy is provided by the time evolution of $B_d\to D_d^+D_d^-$ decays 
and the untagged $B_s\to D_s^+D_s^-$ rate. 

These new strategies may be promising for ``second-generation'' $B$-physics 
experiments, for example LHCb. Their experimental feasibility strongly
depends on the size of the penguin effects in 
$B_{s(d)}\to J/\psi\, K_{\rm S}$ and $B_{d (s)}\to D^{\,+}_{d(s)}\, 
D^{\,-}_{d(s)}$, which are very difficult to calculate and require 
further theoretical studies. Recent experimental results of the CLEO
collaboration on certain non-leptonic $B$ decays, which are dominated by 
penguin contributions, have shown that these topologies may well lead 
to surprises.

\vspace{0.5cm}

\noindent
{\it Acknowledgement}

\vspace{0.1cm}

\noindent
I would like to thank Patricia Ball, Thomas Mannel and Joaquim Matias for 
interesting discussions.


\begin{thebibliography}{99}

\bibitem{bisa}A.B. Carter and A.I. Sanda, {\it Phys.\ Rev.\ Lett.}~{\bf 45}
(1980) 952; {\it Phys.\ Rev.}~{\bf D23} (1981) 1567; I.I. Bigi and 
A.I. Sanda, {\it Nucl.\ Phys.}~{\bf B193} (1981) 85.

\bibitem{ut}L.L. Chau and W.-Y. Keung, {\it Phys.\ Rev.\ Lett.}~{\bf 53} 
(1984) 1802; C. Jarlskog and R.~Stora, {\it Phys.\ Lett.}~{\bf B208} (1988)
268.
 
\bibitem{ckm}N. Cabibbo, {\it Phys.\ Rev.\ Lett.}~{\bf 10} (1963) 531;
M. Kobayashi and T. Maskawa, {\it Progr.\ Theor.\ Phys.}~{\bf 49} (1973)
652.
 
\bibitem{exp}OPAL collaboration (K. Ackerstaff {\it et al.}), {\it Eur.\ 
Phys.\ J.}~{\bf C5} (1998) 379; CDF Collaboration (F. Abe {\it et al.}),
{\it Phys.\ Rev.\ Lett.}~{\bf 81} (1998) 5513; for an updated analysis,
see preprint CDF/PUB/BOTTOM/CDF/4855.

\bibitem{revs}For reviews, see, for instance, {\it The BaBar Physics Book},
eds.\ P.F. Harrison and H.R.~Quinn (SLAC report 504, October 1998); Y. Nir, 
published in the Proceedings of the 18th International Symposium on 
Lepton--Photon Interactions (LP '97), Hamburg, Germany, 28 July--1 August 
1997, eds.\  A. De Roeck and A. Wagner (World Scientific, 
Singapore, 1998), p.\ 295 [hep-ph/9709301]; M. Gronau, {\it Nucl.\ Phys.\
Proc.\ Suppl.}~{\bf 65} (1998) 245; R. Fleischer, 
{\it Int.\ J. Mod.\ Phys.}~{\bf A12} (1997) 2459.

\bibitem{DGamma}For a recent calculation of $\Delta\Gamma_s$, see 
M. Beneke, G. Buchalla, C. Greub, A. Lenz and U. Nierste, preprint 
CERN-TH/98-261 (1998) [hep-ph/9808385].
 
\bibitem{dunietz}I. Dunietz, {\it Phys.\ Rev.}~{\bf D52} (1995) 3048;
R. Fleischer and I. Dunietz, {\it Phys.\ Rev.}~{\bf D55} (1997) 259.

\bibitem{Bsclean}M. Gronau and D. London, {\it Phys.\ Lett.}~{\bf B253}
(1991) 483; R. Aleksan, I. Dunietz and B. Kayser, {\it Z. Phys.}~{\bf
C54} (1992) 653; R. Fleischer and I. Dunietz, {\it Phys.\ Lett.}~{\bf B387}
(1996) 361.

\bibitem{wolf}L. Wolfenstein, {\it Phys.\ Rev.\ Lett.}~{\bf 51} (1983) 1945. 
 
\bibitem{blo}A.J. Buras, M.E. Lautenbacher and G. Ostermaier, {\it Phys.\
Rev.}~{\bf D50} (1994) 3433.

\bibitem{bfm}A.J. Buras, R. Fleischer and T. Mannel, {\it Nucl.\ 
Phys.}~{\bf B533} (1998) 3.

\bibitem{rf-BpiK}R. Fleischer, {\it Eur.\ Phys. J.}~{\bf C6} (1999) 451;
{\it Phys.\ Lett.}~{\bf B435} (1998) 221.

\bibitem{ddf1}A.S. Dighe, I. Dunietz and R. Fleischer, 
{\it Eur.\ Phys. J.}~{\bf C6} (1999) 647.

\bibitem{nirsil}Y. Nir and D. Silverman, {\it Nucl.\ Phys.}~{\bf B345} 
(1990) 301.

\bibitem{PDG}The 1998 Review of Particle Physics, C. Caso {\it et al.},
{\it Eur.\ Phys. J.}~{\bf C3} (1998) 1.

\bibitem{fm1}R. Fleischer and T. Mannel, {\it Phys.\ Lett.}~{\bf B397} (1997)
269.

\bibitem{charles}J. Charles, {\it Phys.\ Rev.}~{\bf D59} (1999) 054007.

\bibitem{BSW}M. Bauer, B. Stech and M. Wirbel, {\it Z. Phys.}~{\bf C29} (1985)
637 and {\bf C34} (1987) 103.

\bibitem{BB}P. Ball and V.M. Braun, {\it Phys.\ Rev.}~{\bf D58} (1998) 
094016.

\bibitem{ambig}See, for example, Y. Grossman and H.R. Quinn, 
{\it Phys.\ Rev.}~{\bf D56} (1997) 7259; J.~Charles, A. Le Yaouanc, 
L. Oliver, O. P\`ene and J.-C. Raynal, {\it Phys.\ Lett.}~{\bf B425} (1998) 
375; A.S. Dighe, I. Dunietz and R. Fleischer, {\it Phys.\ Lett.}~{\bf B433} 
(1998) 147.

\bibitem{neu-ste}M. Neubert and B. Stech, in {\it Heavy Flavours II}, eds.\
A.J. Buras and M. Lindner (World Scientific, Singapore, 1998), p.\ 294--344
[hep-ph/9705292].

\bibitem{HMChiPT1}E. Jenkins and M.J. Savage, {\it Phys.\ Lett.}~{\bf281} 
(1992) 331.

\bibitem{HMChiPT2}B. Grinstein, E. Jenkins, A.V. Manohar, M.J. Savage and
M.B. Wise, {\it Nucl.\ Phys.}~{\bf B380} (1992) 369.

\bibitem{lat1}The UKQCD Collaboration, L. Lellouch {\it et al.}, preprint 
CPT-98-PE-3689 [hep-lat/9809018].

\end{thebibliography}
\end{document}